\documentclass[sn-mathphys,Numbered]{sn-jnl}

\usepackage{graphicx}%
\usepackage{multirow}%
\usepackage{amsmath,amssymb,amsfonts}%
\usepackage{mathrsfs}%
\usepackage[title]{appendix}%
\usepackage{xcolor}%
\usepackage{textcomp}%
\usepackage{manyfoot}%
\usepackage{booktabs}%
\usepackage{algorithm}%
\usepackage{algorithmicx}%
\usepackage{algpseudocode}%
\usepackage{listings}%

\newcommand*{\3}{$^3$He}

\begin{document}

\title[Article Title]{Drag on Cylinders Moving in Superfluid \3-B as the Dimension Spans the Coherence Length}


\author[1]{\fnm{S.} \sur{Autti}}

\author[1]{\fnm{R.P.} \sur{Haley}}

\author[2]{\fnm{A.} \sur{Jennings}}

\author[1]{\fnm{G.R.} \sur{Pickett}}

\author[3]{\fnm{E.V.} \sur{Surovtsev}}

\author[1]{\fnm{V.} \sur{Tsepelin}}

\author*[1]{\fnm{D.E.} \sur{Zmeev}}\email{d.zmeev@lancaster.ac.uk}

\affil[1]{\orgdiv{Department of Physics}, \orgname{Lancaster University}, \orgaddress{ \city{Lancaster}, \postcode{LA1 4YB}, \country{UK}}}

\affil[2]{\orgdiv{RIKEN Center for Quantum Computing}, \orgname{RIKEN}, \orgaddress{ \city{Wako}, \postcode{351-0198},  \country{Japan}}}

\affil[3]{\orgdiv{Kapitza Institute for Physical Problems}, \orgname{Russian Academy of Sciences}, \orgaddress{ \city{Moscow}, \postcode{119334},  \country{Russia}}}


\abstract{Vibrating probes when immersed in a fluid can provide powerful tools for characterising the surrounding medium. In superfluid \3-B, a condensate of Cooper pairs, the dissipation arising from the scattering of quasiparticle excitations from a mechanical oscillator provides the basis of extremely sensitive thermometry and bolometry at sub-millikelvin temperatures. The unique properties of the Andreev reflection process in this condensate also assist by providing a significantly enhanced dissipation. While existing models for such damping on an oscillating cylinder have been  verified experimentally, they are  valid only for flows with scales much greater than the coherence length of \3, which is of the order of a hundred nanometres. With our increasing proficiency in fabricating nano-sized oscillators, which can be readily used in this superfluid, there is a pressing need for the development of new models that account for the modification of the flow around these smaller oscillators. Here we report preliminary results on measurements of the damping in superfluid \3-B of a range of cylindrical nano-sized oscillators with radii comparable to the coherence length and outline a model for calculating the associated drag.}


\maketitle

\section{Introduction}\label{sec1}

In this article, we present measurements on the drag force on nano-scale cylinders moving in superfluid \3-B where the range of cylinder dimensions spans the superfluid coherence length.   This is very appropriate to a volume dedicated to the memory of A.~F.~Andreev as the motion of objects in superfluid \3 is one of the best illustrations of Andreev reflection in action since the process is completely dominated by Andreev reflection.  

In effect, the Andreev reflection behaves as a Maxwell Demon which chooses which quasiparticle excitations can reach the surface of the moving wire and thus interact with the wire via a normal process.  Excitations which are deflected by Andreev processes by the flow fields near the wire and thus cannot impact the wire surface can essentially exchange zero momentum with the wire and thus do not contribute to the wire damping.   Since these processes largely involve quasiholes approaching the wire from the front and quasiparticles approaching the wire from the rear, they introduce an enormous asymmetry in the scattering of holes and particles leading to drag forces an order of magnitude larger than would be expected for particle-hole symmetry.  This magnification of the drag forces means that moving objects in the superfluid can provide incredibly sensitive mechanical quasiparticle detectors (or in other words thermometers) in a system of excitations which in a normal context would be regarded as a high vacuum.   

As a result, such oscillators provide the ``Swiss Army Knife'' of superfluid \3 sensors providing us with quasiparticle detectors (and thus thermometers), quasiparticle generators, sensitive heaters and many more.  With this in mind, we undertook the current research project to extend our knowledge of these sensors to smaller and smaller scales.

 In \3-B at low temperatures, $T <100 \,\mu$K, the normal fluid component can be represented as a gas of ballistic quasiparticle excitations. Under these circumstances, Andreev reflection is responsible for enhanced drag on an object moving in the superfluid. A short review \cite{pickett2014superfluid} summarizes the manifestations of Andreev reflection in superfluid \3 and review \cite{bradley2017andreev} is centred upon the interaction of quasiparticles  with the flow generated by quantum vortices in \3-B. Here we outline the main features necessary for the understanding of the enhanced drag.

 The essence of the effect is that the flow field, set up by the moving object, leads to an increase in the flow velocity over a region of the order of the dimension of the moving object, which skews the dispersion curve of the excitations. A quasiparticle with a low enough energy and momentum $p_F$ directed along the object's momentum, i.e. a quasiparticle approaching the moving object from the rear experiences a potential barrier arising from this enhanced flow and thus is forced to undergo Andreev reflection, the quasiparticle being converted into a quasihole during the process which results in only a negligible exchange of momentum with the object, on the order of $p_F \frac{\sqrt{T\Delta}}{E_F}\sim 10^{-3}p_F$ (we assume energy units of temperature $T$, $\Delta$ and $E_F$ are the superfluid gap and the Fermi energy, respectively). On the other hand, a quasiparticle with the same energy moving in the opposite direction, i.e. approaching the moving object from the front, sees no such potential barrier and can reach the surface of the object to be reflected normally, imparting almost $2p_F$ momentum change on the object.   Quasiholes behave in a similar way, but with their roles reversed: quasiholes can be normally reflected from the rear but are Andreev reflected when approaching the front. This asymmetry leads to a drag force enormously enhanced compared with what would be expected from the normal reflection of all quasiparticles and holes \cite{fisher1989beyond}. 
 
 For a cylinder moving in superfluid \3 in the limit of low velocity $u \ll T / p_F \sim 1$\,mm\,s$^{-1}$ for $T \sim 100\,\mu$K, the drag force per unit length $F$, normalised by the diameter of the cylinder $2R$ and its velocity $u$ is given by \cite{EnricoMichaelPaul1995Besp,Enrico1995}
\begin{equation}\label{Force_Enrico}
    F'=\frac{F}{2Ru}=\frac{\pi}{4} p_F^2 v_F N_F \exp\left({-\frac{\Delta}{T}}\right) 
\end{equation}
Here $v_F$ is the Fermi velocity, and $N_F$ is the density of states at the Fermi level. 

The measured resonance width $\Delta f$ of a resonator can be converted into a drag force using 
\begin{equation}\label{harmonic}
    \frac{F}{u}= 2 \pi\, m\, \Delta f.
\end{equation}
Here $m$ is the resonator mass, which we take to be the mass per unit length. Note that the force in Eq.~(\ref{Force_Enrico}) is the drag force, while that in Eq.~(\ref{harmonic}) is the driving force. These are equal in a dynamic steady state for a pointlike resonator. For a wire resonator, three correction factors need to be applied in these equations \cite{Schmoranzer2019}: the mass becomes an effective mass that describes how the resonator stores energy for a given $u$ (which is taken to be the velocity of the fastest moving part of the resonator), and the two forces become effective as well, describing how the drive inputs power and how the drag force dissipates it. As described below, in this paper we assume that all these correction terms can be neglected. Thus, one can  equate the left-hand side of Eq.~(\ref{Force_Enrico}) and the right-hand side of Eq.~(\ref{harmonic}) normalised by the known diameter of the cylinder.

As one can see, the drag changes very rapidly with temperature due to the Boltzmann factor. This change translates into extreme sensitivity of the damping of an oscillator immersed in \3 to temperature (as much as 20\,\% change in damping per microkelvin at 100\,$\mu$K at saturated vapour pressure).  This sensitivity allows the mechanical oscillators in superfluid \3 to serve as detectors of the density of quasiparticles. Consequently, they can be used as thermometers \cite{bauerle1998temperature} as well as the sensing elements in ultra-sensitive bolometers \cite{bradley1995potential, autti2023quest}.

 By reducing the cross-section of the oscillating cylinder, one can further boost the sensitivity of the probe to the changes in the environment \cite{bradley2017operating}. However, in \3 the coherence length, characterising the size of the Cooper pair, is relatively large, i.e. tens of nanometers \cite{vollhardt1990superfluid}. The order parameter of a $p$-wave superfluid is also suppressed near the surface of a macroscopic object immersed in \3 with the recovery to the bulk value occurring over a distance of the order of the coherence length $\xi_0$. If the object is nano-sized, i.e. of similar size to the coherence length, then the potential barriers seen by approaching excitations will not be so high because of the gap suppression. As a result, the Andreev reflection of quasiparticles will not be as effective in enhancing the drag. Moreover, the effect can be observable on the quasi-macroscopic objects, whose size is several coherence lengths, due to the smallness of the  parameter $up_F/\Delta$. Decreasing  the radius further would lead to a situation when the coherence length is much bigger than the radius of the wire. In such a case there is no gap suppression at the surface  of the object and the energy of the condensate is only slightly perturbed. For the same reason, the flow field around the object is also suppressed. And that is the case of `pure' scattering of quasiparticles off the object which is for instance realised in the experiments with negative ions moving in superfluid \3.

\section{Two contributions to the force}

To make a comparison between the extremal behaviours for objects greater and smaller than the coherence length we take a mixed approach, where we effectively assume that the coherence length is zero, but take into account the flow field suppression for the smaller objects, making sense of earlier calculations for the motion of ions in the superfluid. 

In order to estimate the above-mentioned enhancement of the drag force due to Andreev reflection we start from the standard expression for the force $\mathbf{F}$, acting on an object moving with velocity $\mathbf{u}$ through a gas of quasiparticles \cite{baym1979mobility}:
 \begin{eqnarray}
 \label{Force_0}
 \mathbf{F}=\frac{d\mathbf{P}}{dt}=-\sum_{{\mathbf{p},\mathbf{p}^{'}}}(\mathbf{p}^{'}-\mathbf{p})n_{\mathbf{p}}(1-n_{\mathbf{p}^{'}})\frac{2\pi}{\hbar}\delta(E_{\mathbf{p}}^{'}-E_{\mathbf{p}}-\mathbf{u}(\mathbf{p}^{'}-\mathbf{p}))|t(\mathbf{p}\rightarrow\mathbf{p}^{'},\mathbf{u})|^2,    
 \end{eqnarray}
where $\mathbf{p}$, $\mathbf{p}^{'}$ are the momenta of incident and reflected quasiparticles respectively, $n_{\mathbf{p}^{'}}$ -- the Fermi-distribution of quasiparticles in the laboratory frame of reference, $E_{\mathbf{p}}$ -- the energy of Bogolyubov quasiparticle with momentum $\mathbf{p}$ in the rest frame of liquid, the delta-function incorporates the condition that the collision is elastic in the rest frame of the object. The most important information for us is hidden in the dependence of the amplitude of  quasiparticle scattering from the momentum $\mathbf{p}$ to momentum $\mathbf{p}^{'}$ on the velocity of the object: $t(\mathbf{p}\rightarrow\mathbf{p}^{'},\mathbf{u})$. One can separate the effect without Andreev reflection by the following transformations. Suppose the object and the liquid are at rest. Then we can write down the condition of equilibrium in the form:
\begin{eqnarray}
\label{Equil}
 \frac{d\mathbf{P}}{dt}=-\sum_{{\mathbf{p},\mathbf{p}^{'}}}(\mathbf{p}^{'}-\mathbf{p})n_{\mathbf{p}}(1-n_{\mathbf{p}^{'}})\frac{2\pi}{\hbar}\delta(E_{\mathbf{p}}^{'}-E_{\mathbf{p}})|t(\mathbf{p}\rightarrow\mathbf{p}^{'},\mathbf{u}=0)|^2=0.    
 \end{eqnarray}
 Subtracting Eq.~\ref{Equil} from Eq.~\ref{Force_0} we arrive at an expression with two summations: 
\begin{align}
 \label{Force_two_parts}
&\mathbf{F}&=~&-\sum_{{\mathbf{p},\mathbf{p}^{'}}}\biggl((\mathbf{p}^{'}-\mathbf{p})n_{\mathbf{p}}(1-n_{\mathbf{p}^{'}})\frac{2\pi}{\hbar}\left|t(\mathbf{p}\rightarrow\mathbf{p}^{'},\mathbf{u}=0)\right|^2\times \nonumber\\
& & &\times\Bigl(\delta(E_{\mathbf{p}}^{'}-E_{\mathbf{p}}-\mathbf{u}(\mathbf{p}^{'}-\mathbf{p}))-\delta(E_{\mathbf{p}}^{'}-E_\mathbf{p})\Bigr)\biggr)-\\
& & & -\sum_{{\mathbf{p},\mathbf{p}^{'}}}\biggl((\mathbf{p}^{'}-\mathbf{p})n_{\mathbf{p}}(1-n_{\mathbf{p}^{'}})\frac{2\pi}{\hbar}\left[|t(\mathbf{p}\rightarrow\mathbf{p}^{'},\mathbf{u})|^2-|t(\mathbf{p}\rightarrow\mathbf{p}^{'},\mathbf{u}=0)|^2\right]\times \nonumber\\
& & & \times\delta\Bigl(E_{\mathbf{p}}^{'}-E_{\mathbf{p}}-\mathbf{u}(\mathbf{p}^{'}-\mathbf{p})\Bigr)\biggr)\,. \nonumber
\end{align}

 The first sum has been calculated in most works concerning the mobility of ions in the superfluid \3. For a small enough velocity of the object, it can be simplified to the expression:
\begin{eqnarray}
\mathbf{F}_0&=&-\frac{1}{2}\sum_{{\mathbf{p},\mathbf{p}^{'}}}(\mathbf{p}^{'}-\mathbf{p})\frac{\partial n}{\partial E}\mathbf{u}(\mathbf{p}^{'}-\mathbf{p})\frac{2\pi}{\hbar}\delta(E_{\mathbf{p}}^{'}-E_{,\mathbf{p}})|t(\mathbf{p}\rightarrow\mathbf{p}^{'},\mathbf{u}=0)|^2\\
&=&-3\mathbf{u}N_FE_Fp_F\int_{-\infty}^{+\infty}d\eta_p\left|\frac{\eta_p}{E}\right|\left(-\frac{\partial n}{\partial E}\right)\sigma_{tr}(\eta_p),  \nonumber
\end{eqnarray}
 where the substitution is made $n_{\mathbf{p}}-n_{\mathbf{p}^{'}}\approx-\frac{\partial n}{\partial E}\mathbf{u}(\mathbf{p}^{'}-\mathbf{p})$, $\sigma_{tr}(\eta_p)$ --the transport cross-section of the object, $\eta_p\approx\frac{(p-p_F)p_F}{m}$. 
 The second sum describes the enhancement of the force due to Andreev reflection that arises in the presence of the superfluid velocity field:
 \begin{eqnarray}\label{Force_sec_sum}
&\nonumber F_{a}= - \mathbf{F}_{a}\frac{\mathbf{u}}{u}=\\
    &-\sum_{{\mathbf{p},\mathbf{p}^{'}}}\frac{\mathbf{u}}{u}\left((\mathbf{p}^{'}-\mathbf{p})n_{\mathbf{p}}(1-n_{\mathbf{p}^{'}})\frac{2\pi}{\hbar}[|t(\mathbf{p}\rightarrow\mathbf{p}^{'},\mathbf{u})|^2-|t(\mathbf{p}\rightarrow\mathbf{p}^{'},\mathbf{u}=0)|^2]\right.\cdot\\
    &\cdot\left.\delta(E_{\mathbf{p}}^{'}-E_{\mathbf{p}}-\mathbf{u}(\mathbf{p}^{'}-\mathbf{p})\right).\nonumber
\end{eqnarray}
If we assume $T\ll\Delta$, then $1-n_{\mathbf{p}^{'}}\approx 1$ and the Andreev force takes the form:
\begin{eqnarray}
&    F_a\approx-\sum_{{\mathbf{p},\mathbf{p}^{'}}}\frac{\mathbf{u}}{u}\left((\mathbf{p}^{'}-\mathbf{p})n_{\mathbf{p}}\frac{2\pi}{\hbar}|\delta t(\mathbf{p}\rightarrow\mathbf{p}^{'},\mathbf{u})|^2\right.\cdot\\
&    \left.\delta(E_{\mathbf{p}}^{'}-E_{\mathbf{p}})\right),\nonumber
\end{eqnarray}
where $\delta t(\mathbf{p}\rightarrow\mathbf{p}^{'},\mathbf{u})$ is the new function of scattering, dependent on  the velocity of the object. After summation over the directions of incident and outgoing quasiparticles, it can be simplified by introducing another transport cross-section function $\tilde{\sigma}_{tr}(\eta_p,up_F)$ defined by $\delta t$ instead of $t$:
\begin{eqnarray}
\label{cross_sec}
\mathbf{F}_a=-\frac{\mathbf{u}}{u}N_FE_F\int_{-\infty}^{+\infty}d\eta_p \left|\frac{\eta_p}{E}\right|n_{\mathbf{p}}\tilde{\sigma}_{tr}(\eta_p,up_F).
\end{eqnarray}

As it was shown in \cite{fisher1989beyond} the function $\tilde{\sigma}_{tr}(\eta_p,up_F)$ can be approximated by the formula:
\begin{equation}
   \tilde{\sigma}_{tr}(\eta_p,up_F)=\tilde{\sigma}_0(\eta_p)\theta(E(\eta_p)-\Delta -\lambda up_F)\approx\tilde{\sigma}_0(\eta_p)\theta(|\eta_p| -\sqrt{2\lambda up_F\Delta}),
\end{equation}
where $\lambda$ is the geometrical factor which is of the order of unity for macroscopic objects and zero for ions, $\tilde{\sigma}_0(\eta_p)$ describes the magnitude of the contribution to the cross section from the scattering processes in the presence of the flow around the object (i.e. Andreev reflection), $\theta(x-x_0)$ is the Heaviside step function. As it is clear from the definition, parameter $\lambda$ determines the energy cut off on the scale of $up_F$ for the quasiparticles that can reach the object. It has the meaning of an averaged value of the effective potential barrier for the given flow field and thus it depends on the geometry of the object.  By changing $\lambda$ one can interpolate between two regimes: small object without backflow and macroscopic object with backflow. The characteristic size of the object where the interpolation occurs can be estimated from the fact, that the energy of the small object in superfluid \3 is a small part of the condensation energy $N_F\Delta^2$. The small parameter for the point-like object is $\frac{R^2}{\xi_0^2}$ ($R$ -- size of the object) and $\frac{R}{\xi_0}$  is for the long object with length $L\gg\xi_0$ and transverse size $R$. The overall energy is found by integration over the volume where perturbation occurs: $\xi_0^3$ for the point-like object and $L\xi_0^2$ for the long object. Thus, the perturbation energies in considered above cases are the following:  $N_F\Delta^2R^2\xi_0$ and $N_F\Delta^2RL\xi_0$ for the point-like object and long object correspondingly. By contrast, the energy of backflow around a macroscopic object is on the order of $R^3N_F\Delta^2\frac{\xi_0^2}{R^2}$ or $R^2LN_F\Delta^2\frac{\xi_0^2}{R^2}$ depending on the geometry, where the last multiplier comes from the gradient energy of the superfluid in the region near the object. The energy of the backflow and the static energy of the small object are of the same order if $\xi_0\sim R$ in both cases. From this estimation, one can conclude that the parameter $\lambda$ is in fact a function of $\xi_0/R$. A more detailed analysis of this fact is made in section~\ref{sec12}. 

Let us estimate the two contributions to the force in the same limit of low temperatures and under the assumption of small velocities $up_F\ll T$. In this case, it is important to know the behaviour of $\sigma_{tr}(\eta_p)$ and $\tilde{\sigma}_{0}(\eta_p)$ at $\eta_p/\Delta\rightarrow 0$. As follows from the  calculations of Baym {\it et al.}~\cite{baym1979mobility}, $\sigma_{tr}(0)\sim\sigma_{tr}^N\frac{1}{k_F R}$, where $\sigma_{tr}^N\sim \pi R^2$ is the transport cross-section of an ion with radius R in the normal state. For the ions, $k_FR\sim 10$ and for macroscopic objects (in vibrating wire experiments), $k_FR>1000$. Thus, for the latter case, one can assume that $\sigma_{tr}(0)\approx 0$ and the leading term on $\eta_p$ is $\sigma_{tr}(\eta_p) \approx \sigma_{tr}^N\frac{\eta_p^2}{\Delta^2}$. As a result, we arrive at an estimation of the force:
\begin{eqnarray}
    F_0&\sim& uN_FE_Fp_F\int\limits_{-\infty}^{+\infty}d\eta_p\left|\frac{\eta_p}{E}\right|\left(-\frac{\partial n}{\partial E}\right)\sigma_{tr}^N\frac{\eta_p^2}{\Delta^2}\nonumber\\  
   & \approx& 2uN_FE_Fp_F\sigma_{tr}^N\int\limits_{0}^{+\infty}d\eta_p\left(\frac{\eta_p}{\Delta}\right)^3\frac{1}{T}e^{-\frac{\Delta}{T}-\frac{\eta_p^2}{2\Delta T}}
   \\ 
    & =& 4uN_FE_Fp_F\frac{\Delta}{T}e^{-\frac{\Delta}{T}}\sigma_{tr}^N\int\limits_{0}^{+\infty}xe^{-x}dx\sim uN_FE_Fp_F\frac{T}{\Delta}e^{-\frac{\Delta}{T}}\sigma_{tr}^N,\nonumber
\end{eqnarray}
where we used the following approximations $E\approx \Delta+\frac{\eta_p^2}{2\Delta T}$, $-\frac{\partial n}{\partial E}\approx \frac{1}{T}e^{-\frac{E}{T}}$.
Oppositely, for the second contribution to the drag force (as it is claimed in \cite{fisher1989beyond}) the value of $\tilde{\sigma}_0(0)$ is not zero and is estimated as $\tilde{\sigma}_0(0)\sim\sigma_{tr}^N$. Apparently, that is a result of the classical description of the scattering process. As a consequence, the order of the Andreev force for small velocities $up_F\ll T$ is found to be:
\begin{eqnarray}
 \label{Force_a}
   F_a&\sim& N_FE_F\int\limits_{-\sqrt{2\lambda up_F\Delta}}^{\sqrt{2\lambda up_F\Delta}}d\eta_p \left|\frac{\eta_p}{E}\right|n_{\mathbf{p}}{\sigma}_{tr}^N\approx2N_FE_F{\sigma}_{tr}^N\int\limits_{0}^{\sqrt{2\lambda up_F\Delta}}d\eta_p \frac{\eta_p}{\Delta}e^{\frac{-\Delta}{T}}\nonumber\\
   &=&2\lambda uN_FE_Fp_Fe^{-\frac{\Delta}{T}}\sigma_{tr}^N.
\end{eqnarray}

Comparing the two results, one can see that the first term is much smaller in the limit considered and the enhancement of the force for a macroscopic object is of the order of $\frac{\Delta}{T}\sim 10$ for T $\sim200~\mu$K. It should be noted that the enhancement arises only from the reduction of the transport cross-section of the object at small energies in comparison with the effective cross-section entering the expression for the Andreev force. 

The existing experimental data for ions \cite{nummila1989experiments} does not contradict our  estimation  for $F_0$ as well as the data for the wires confirms the dependence of $F_a$ on the parameter $up_F/T$ specific for the effect of Andreev reflection. Nevertheless,  there is no rigorous proof that $\tilde{\sigma}_0(0)$ is of the order of $\sigma_{tr}^N$ at the moment. We can only suggest a qualitative explanation of that fact. It is known that the differential cross-section of fast particles ($k_FR\gg1$) on a sphere could be divided into two parts - the classical and the diffraction one. The second contribution arises from interference of the wave function of the particle behind the object. This part significantly enhances the probability of reflection at small angles. In the superfluid state of \3 the situation is even more complicated. As was found by Baym {\it et al.}~\cite{baym1979mobility} and pointed out by Tsutsumi~\cite{tsutsumi2017scattering}, the contribution of the interference arises also for the  back-scattering processes and makes its probability vanishingly small.  And that is the reason for the mentioned dependence of $\sigma_{tr}(\eta_p)$ at small $\eta_p$. In the presence of Andreev reflection, there is probably no interference between quasiparticles and quasiholes scattered from the object since only one part of them is reflected by normal processes. As a result, the transport cross section may recover its value of the order of $\sigma_{tr}^{N}$.

\section{Results}\label{sec2}

In Lancaster, we have a long experience in using vibrating wire resonators in superfluid \3. In an extension of this work we have recently succeeded in producing circular cross section nanowires with radii less than $10^{-7}$\,m and with lengths of order of a millimetre \cite{autti2023long}. We have tested these nanowire oscillators in superfluid \3, since one of the aims of this series of experiments was to observe the behaviour when the diameter of an oscillating wire became comparable to the coherence length.

\begin{figure}
    \centering
    \includegraphics[width=0.9\linewidth]{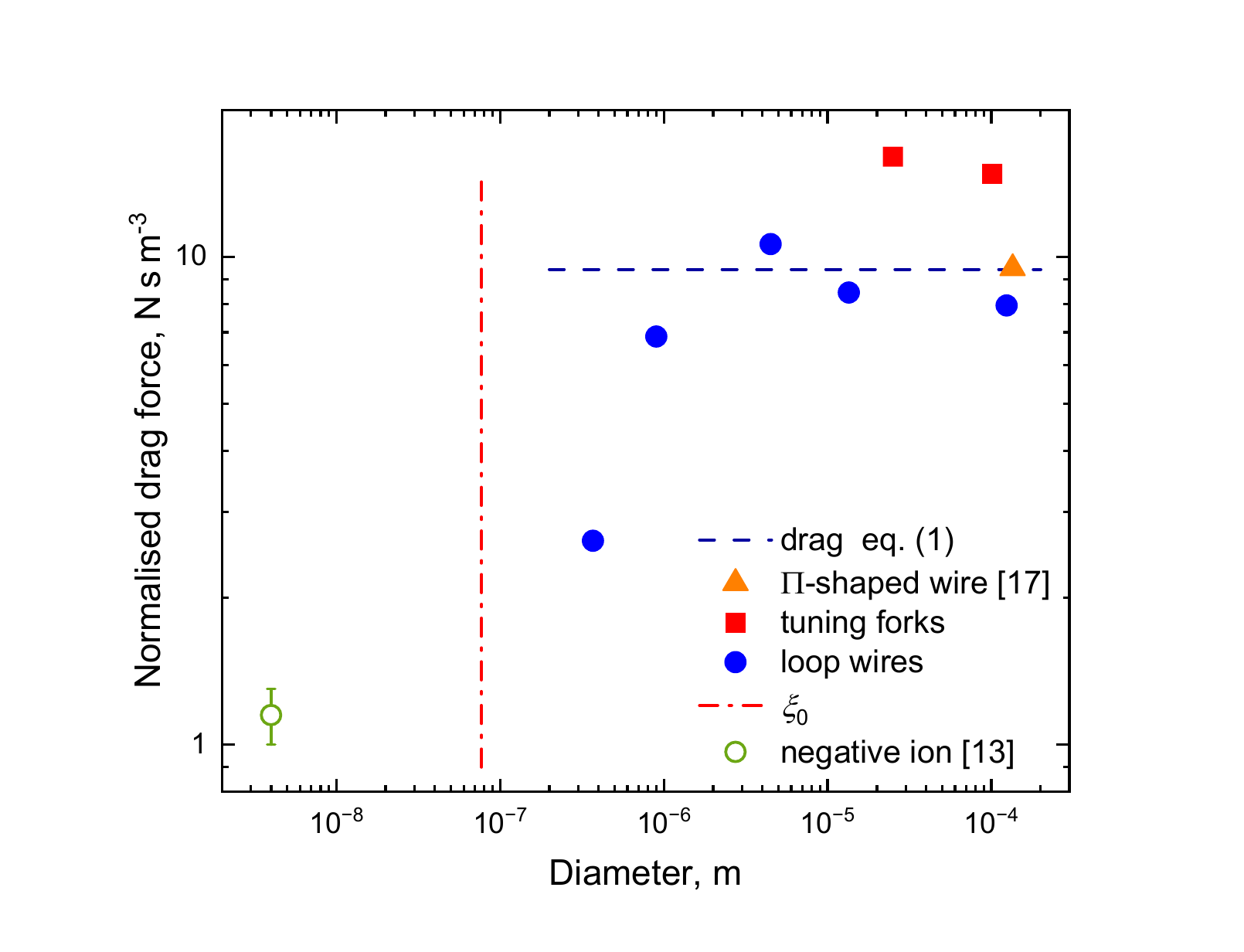}
    \caption{The normalised drag force $F'$ of Eq.~\ref{Force_Enrico} for probes of various sizes at 220\,$\mu$K at the saturated vapour pressure. The  prediction given by Eq.~\ref{Force_Enrico} for this temperature is shown by a horizontal dashed line. The probes include: quartz tuning forks (squares), a $\Pi$-shaped wire~\cite{bradley2016breaking} (triangle) and wires bent in the shape of a semi-circular loop (filled circles). For the quartz tuning forks we took the dimension of the fork normal to its motion as $2R$ in Eq.~\ref{Force_Enrico}. The drag force on a negative ion (open circle) has been extrapolated from mobility data in \cite{nummila1989experiments}.  The vertical dash-dotted line shows the length scale for the coherence length $\xi_0$. Intrinsic dissipation in all mechanical probes is negligible at this temperature.}
    \label{F_vs_D}
\end{figure}

In our experiments we measured the drag force on magneto-motively excited wire resonators vibrating in \3. The width of the measured resonance is proportional to the drag force and the ratio of the width of the resonance and the diameter of the wire should remain constant in the $R\gg \xi_0$ limit according to Eq.~\ref{Force_Enrico}. This is indeed the case for wires with diameters greater than 1\,$\mu$m or so. However, for smaller wires the measured damping force is dramatically smaller than that expected from Eq.~\ref{Force_Enrico}.  The results \cite{Jennings2022} are shown in Fig.~\ref{F_vs_D} where we can see that the effective damping cross section of the wires decreases rapidly as the wire diameter approaches $\xi_0$. 

We note that the damping force shown in Fig.~\ref{F_vs_D} is calculated using Eq.~(\ref{harmonic}), that is, for a rigid cylindrical rod moving normally to its axis, while in reality most of our vibrating wires were bent in a semi-circular shape. To obtain the drag force, one should adjust the conversion from resonance width to drag force by a factor describing how the resonator stores energy (effective mass) and how the energy is dissipated (effective resonance width)~\cite{Schmoranzer2019}. Also, the driving force needs similar adjustment. Together these make a factor of the order of unity that depends on the resonator geometry. The scatter of the wires data above diameter $10^{-6}$\,m can be attributed to the differences in the resonator geometries. In particular, the thickest wire in this study was bent in the shape of the letter $\Pi$~\cite{bradley2016breaking,autti2020fundamental,autti2023transport}, (triangle in Fig.~\ref{F_vs_D}). We note that even data for two quartz tuning forks~\cite{blaauwgeers2007quartz} of different dimensions with geometries that radically differ from the geometry of a circular cylinder is not dissimilar to the data for circular wires (squares in Fig.~\ref{F_vs_D}). We used the thicknesses of the forks as a substitution for the diameter $2R$ in Eq.~\ref{Force_Enrico}. Thus, we argue that the geometry factors can be neglected for the cylindrical wires.

We observed similar ratios of the damping force for all our probes down to the lowest temperatures of 150\,$\mu$K as well. We chose a temperature of 220\,$\mu$K to illustrate the data in Fig.~~\ref{F_vs_D} as at this temperature the damping on the thickest probes is sufficiently greater than their intrinsic damping in vacuum, while at the same time the thinnest probe is not overdamped.

\section{Discussion}\label{sec12}

\subsection{Modification of effective potential of quasiparticles  around a thin cylinder}
\begin{figure}
    \centering
    \includegraphics[width=0.7\linewidth]{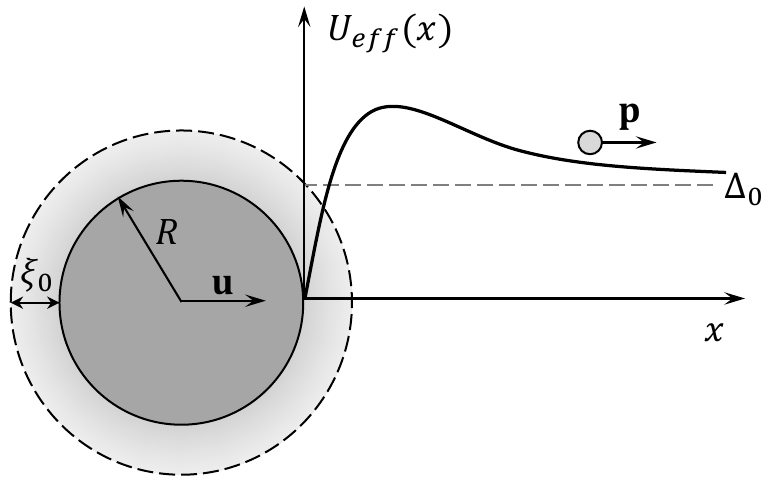}
   \caption{Cylindrical wire of radius $R$ is moving in the $x$-direction with velocity $\mathbf{u}$. The order parameter of the B phase is suppressed in the region of thickness on the order of $\xi_0$ near the wire surface. A quasihole with momentum $\bf{p}\parallel\hat{x}$ is moving towards the wire along the $x$-axis in the presence of the effective potential $U_{eff}(x)$ which consists of two contributions: potential flow field around the wire and spatial dependence of quasiparticle energy gap $\Delta(x)$ due to the order parameter suppression near the wire surface. If the energy of the quasihole is less than the height of the barrier then it reflects by means of the Andreev process with practically no momentum change. The height of the barrier decreases with increasing ratio $\xi_0/R$, Eq.~(\ref{U_max}).}
    \label{Geometry}
\end{figure}

As we pointed out in the previous section, the Andreev force for small velocities depends linearly on the parameter $\lambda$, which is of the order of unity for a macroscopic object (geometric factor). This parameter describes the cutoff energy (in the units of $up_F$) for incident quasiparticles that can `hit' the object and significantly change its momentum (Fig.~\ref{Geometry}). For example, for a cylinder of radius $R$ quasiparticles with momentum parallel to the velocity of the object ($\mathbf{p}\mathbf{u}>0$) gain additional positive potential energy (counted from the level at infinity) due to superflow around the object:
\begin{eqnarray}
    U(\mathbf{r},\mathbf{\hat{p}})=up_F\left(2(\mathbf{n}\mathbf{\hat{p}})(\mathbf{n}\mathbf{\hat{u}})-(\mathbf{\hat{p}}\mathbf{\hat{u}})\right)\frac{R^2}{r^2},
\end{eqnarray}
where $\mathbf{r}$ is the coordinate in the polar frame, $\mathbf{n}$ -- unit vector of normal to the surface, $\mathbf{\hat{u}}$, $\mathbf{\hat{p}}$ -- unit vectors in the direction of $\mathbf{{u}}$ and $\mathbf{{p}}$ correspondingly. As one can see the maximum height of the potential barrier corresponds to $\mathbf{n}\parallel \mathbf{u}$, $\mathbf{p}\parallel \mathbf{u}$  and equals $up_F$. Since the change of momentum for specular reflection is equal to $2(\mathbf{n}\mathbf{p})\mathbf{n}$, the maximum of the momentum change in the direction of the object velocity is also attained for $\mathbf{n}\parallel \mathbf{u}$. We will use this fact for the further estimation. 

\begin{figure}
    \centering
    \includegraphics[width=0.8\linewidth]{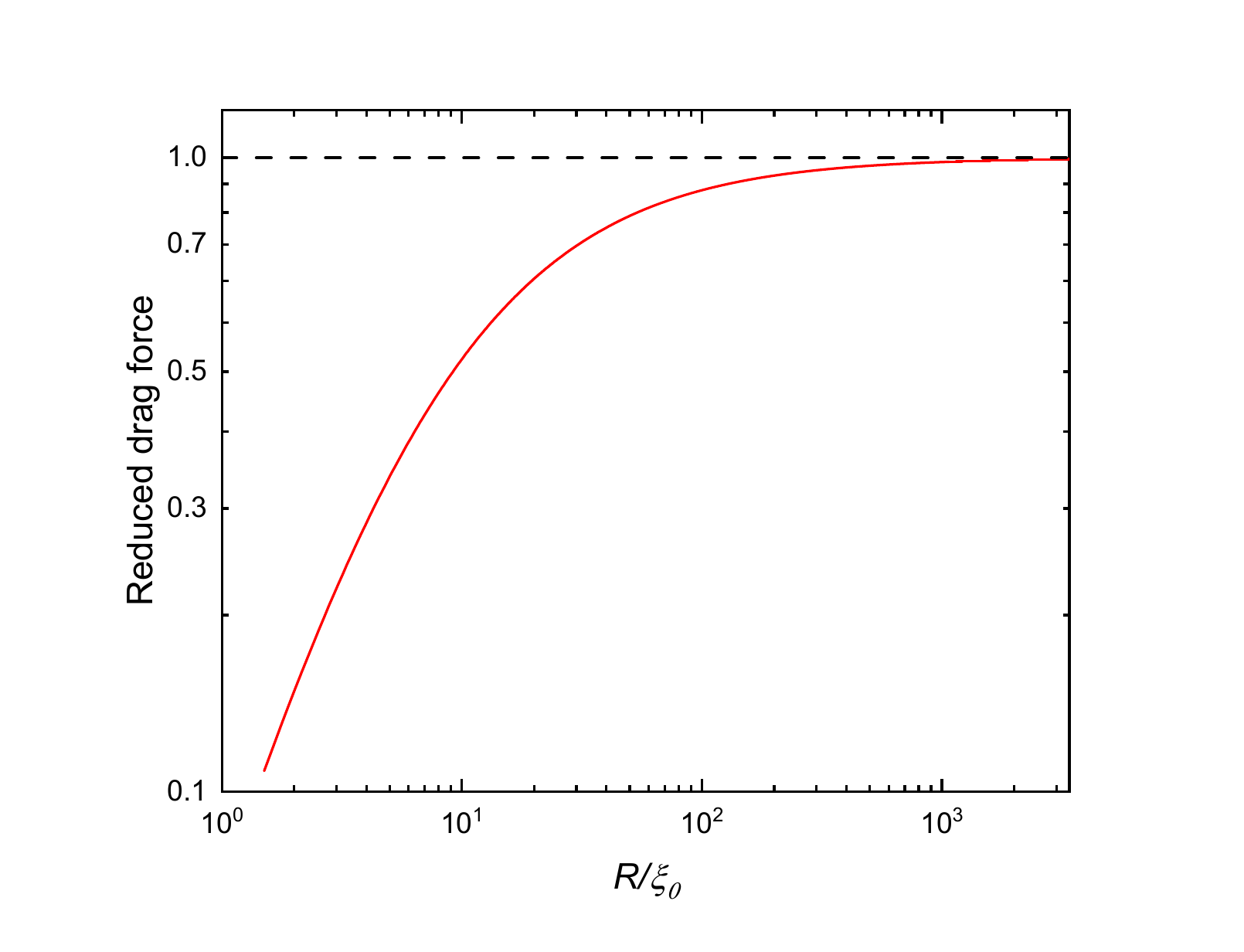}
   \caption{Estimation of the Andreev contribution to the drag force as a function of $R/\xi_0$. The force is normalised to its value recovered in the $R\gg\xi_0$ limit.}
    \label{F_theor}
\end{figure}

There is another effect that influences the Andreev reflection, namely the order parameter suppression near the surface of the cylinder. This effect arises due to quasiparticle scattering 
from the surface. Hereinafter we will assume that the order parameter suppression does not depend on the flow, (this is true for sufficiently small velocities of the wire and in the region $|R-r|>\xi_0$). For the case of the cylindrical wire, the order parameter of the B phase takes the form:
\begin{equation}
   A_{\mu j}(\mathbf{r})=e^{i\varphi}R^S_{\mu\nu}\left(\Delta_{\perp}(\mathbf{r})n_{\nu}n_j+\Delta_{\parallel}(\mathbf{r})[\delta_{\nu j}-n_{\nu}n_j]\right), 
\end{equation}
where $\Delta_{\parallel}$, $\Delta_{\perp}$ are the amplitudes of the order parameter perpendicular or parallel to the vector $\mathbf{n}$ correspondingly, $R_{\mu\nu}^S$ -- an orthogonal matrix in spin space. In the region far from the wire surface $|\mathbf{r}-R|\gg\xi_0$ the order parameter of the B phase becomes isotropic, i.e. $\Delta_{\parallel}=\Delta_{\perp}$. The values of the $\Delta_{\perp,\parallel}$ at the surface depend on the type of quasiparticle scattering on it, which has two limits -- specular and diffusive. In the first case $\Delta_{\parallel}$ is not suppressed at all, while $\Delta_{\perp}$ goes to zero at the surface \cite{Ambega}. The $p$-dependence of quasiparticle gap energy is defined by the expression:
\begin{eqnarray}
    |\Delta(\mathbf{\hat{p}},\mathbf{r})|^2=A_{\mu i}(\mathbf{r})A_{\mu j}^{*}(\mathbf{r})\hat{p}_i\hat{p}_j=\Delta_{\parallel}^2(\mathbf{r})+[\Delta_{\perp}^2(\mathbf{r})-\Delta_{\parallel}^2(\mathbf{r})](n_i\hat{p}_i)^2,
\end{eqnarray}
where the property of orthogonal matrix $R_{\mu\nu}^SR_{\mu\varphi}^S=\delta_{\nu\varphi}$ was used.
In what follows we use the assumption of specular boundary conditions for simplicity. For distances $|r-R|>\xi_0$ the change of $\Delta_{\perp}$ is small and we arrive at the effective potential contribution: 
\begin{eqnarray}
\Delta(\mathbf{\hat{p}},\mathbf{r})\approx\Delta_0+(\Delta_{\perp}(\mathbf{r})-\Delta_0)(n_i\hat{p}_i)^2,
\end{eqnarray}
where we use the notation $\Delta_{\parallel}(r)=\Delta_0=\mathrm{const}$.
Now we can combine two potentials and find the resulting effective potential for quasiparticles:
\begin{eqnarray}
\label{U_eff_1}
    U_{eff}(\mathbf{r},\mathbf{p})=up_F\left(2(\mathbf{n}\mathbf{\hat{p}})(\mathbf{n}\mathbf{\hat{u}})-(\mathbf{\hat{p}}\mathbf{\hat{u}})\right)\frac{R^2}{r^2}-(\Delta_0-\Delta_{\perp}(\mathbf{r}))(\mathbf{n}\mathbf{\hat{p}})^2,~|r-R|\gg\xi_0,
\end{eqnarray}
where the energy is counted from its value at infinity. The next simplification is the use of the fact, that the minimum value of $2R/\xi_0$ used in the experiments with wires is  approximately $5$. Thus one can apply 1D approximation for the dependence of $\Delta_{\perp}(r)$, i.e. $\Delta_{\perp}(\mathbf{r})-\Delta_0\approx -a\Delta_0e^{-\frac{|r-R|}{\xi_{\perp}}}$, where $a$ is a positive constant of the order of unity and $\xi_{\perp}\sim \xi_0$. As it is clear from the above expression the first term in Eq.~(\ref{U_eff_1}) has a negative derivative while the second has a positive one. Therefore the maximum of potential $U_{eff}(r)$ for the given direction of $\mathbf{p}$, which is responsible for the value of the cutoff energy, will decrease. As an estimation of the reduction of $\lambda$, let us consider the direction of maximum momentum change, i.e. $\mathbf{p}\parallel\mathbf{u}$. In the limit $R\gg\xi_0$ the effect of the finite $\xi_0$ is estimated with logarithmic accuracy as follows:
\begin{eqnarray}
\label{U_max}
    U_{eff}^{max}\left(\frac{\xi_0}{R}\right)\approx \frac{U_{eff}^{max}(0)}{\left(1-\alpha\frac{\xi_0}{R}\ln{\left(2\frac{up_F}{\Delta}\frac{\xi_0}{R}\right)}\right)^2},
\end{eqnarray}
where $\alpha$ is defined as $\xi_{\perp}=\alpha\xi_0$. 
Since we assume that parameter $\lambda$ is proportional to the height of the barrier then we can use the above expression for the estimation of the reduction of the Andreev contribution to the drag force. For instance, if we take the parameters from the experiment discussed earlier, in particular $R/\xi_0\sim 2.5$, $up_F/\Delta\sim 0.01$, $\alpha=\frac{1}{\sqrt{2}}$ (as it follows from Ginzburg-Landau approximation $\xi_{\perp}(\tau)=\frac{1}{\sqrt{2}}\xi(\tau)$, where $\tau=1-\frac{T}{T_c}$, $T_c$ -- superfluid transition temperature, and $\lim\limits_{\tau\rightarrow 1} \xi(\tau)=\xi_0$), then  $\lambda$ is decreased by a factor of $10$. The drag force estimated in this way as a function of $R/\xi_0$ is presented in Fig.~\ref{F_theor}. The curve qualitatively follows the experimental data in Fig.~\ref{F_vs_D}.

\subsection{Andreevless limit}

The thinnest wire used in this study had a diameter of $ 4\times 10^{-7}$\,m which is about five times larger than the coherence length. In the limit $R\ll \xi_0$ one would expect the cylinder to act as an impurity and not enhance flow or distort the superfluid gap \cite{Kuorelahti2018}, resulting in no Andreev process taking place. This limit can be realised in experiments with negative ions \cite{nummila1989experiments}. The normalised drag force, recalculated from mobility data measured in these experiments and extrapolated to 220\,$\mu$K, is shown in Fig.~\ref{F_vs_D}.  Here the damping is much lower, as expected.
 
We note that the negative ion mobility measured by Ikegami {\it et al.}~\cite{ikegami2013mobility} near the free surface of \3-B is significantly smaller than that measured in bulk. However, the mobility of negative ions in the vicinity of the free surface is significantly reduced by interactions with surface-bound excitations \cite{tsutsumi2017scattering}.

\subsection{Implications for critical velocity and supercritical motion}

For a negative ion moving uniformly, the observed critical velocity is the Landau velocity $v_L=\Delta/p_F$ \cite{nummila1989experiments}, as opposed to $v_c =\frac{1}{3}v_L$ predicted and observed for a thick cylinder oscillating in \3-B \cite{Lambert1992,carney1989extreme}. 
The size dependence of the critical velocity is an intriguing question. In the case of \3-B this can be explored by tuning the pressure, and therefore $\xi_0$, giving an opportunity to scan the $R/\xi_0$ ratio over half an order of magnitude for the same wire. 

A thick wire moving in \3-B uniformly \cite{zmeev2014method} at velocities above $v_c$ does not show dramatic dissipation~\cite{bradley2016breaking}. Supercritical motion of such a wire allows to reveal dynamics of Andreev-bound surface excitations \cite{autti2020fundamental, autti2023transport}. Here again, the behaviour is dramatically different, with the mobility of ions moving uniformly  significantly reduced above $v_L$~\cite{ahonen1976mobility,nummila1989experiments}. Theoretical description of such cross-over is an open problem and needs a separate approach, e.g.~\cite{Kuorelahti2018,Andreev2023}.

\section{Conclusion}\label{sec13}
Vibrating wire resonators have been important tools for probing the behaviour of excitations in the superfluid.  Here the question of resolution arises. It would be advantageous to maintain quasiparticle sensitivity down to the smallest size to which we can make the transducers. However, the present work makes it clear, as is expected, that the coherence length $\xi_0$ defines the lower limit of resolution for this method of quasiparticle detection. 

On a more fundamental level, nano-sized vibrating wire resonators should enable scanning a cross over from  superflow past a single impurity to superflow past a solid wall in the foreseeable future. 

\backmatter

\section{Dedication to A.F. Andreev} 

EVS and DEZ were students of Alexander Fedorovich and we have retained fond memories of A.F. He has been a great influence on our scientific lives and careers and an inspiration to look up to. Despite his demanding bureaucratic roles as the Director of Kapitza Institute and Vice-President of the Russian Academy of Sciences, he still found time to deliver an exceptional course on low-temperature physics to all Masters students at the Kapitza Institute. The seminars at the Institute chaired by A.F. were an invaluable source of education, both in physics and in scientific discussion. Finally, Andreev's devotion to the Institute, its employees and students is difficult to overestimate.

GRP, and the Lancaster Ultralow Temperature Group, began collaborating with the Kapitza Institute during the time of easier access arriving with the ending of the Soviet Union.  Co-operation subsequently blossomed with visits in both directions by Lancaster and Kapitza Institute workers, notably by several Kapitza Institute directors, Alexandr Fedorovich himself, V-A. S. Borovik-Romanov and V. V. Dmitriev and a vibrant collaboration ensued which has continued to this day. 

\section{Data availability}
The manuscript has associated data in a data repository https://doi.org/10.17635/lancaster/researchdata/669.

\section{Acknowledgements}

This work was funded by UKRI (Grant Nos. EP/P024203/1, EP/W015730/1, ST/T006773/1) and EU H2020 European Microkelvin Platform (Grant Agreement 824109).

\bibliography{sn-bibliography}

\end{document}